\documentclass[english,aps,pra,reprint,noshowpacs,superscriptaddress]{revtex4-1}   
\usepackage[T1]{fontenc}	
\usepackage[utf8]{inputenc}	
\usepackage{geometry}		
\usepackage{graphicx}
\usepackage[above,below]{placeins}	
\usepackage{times}
\usepackage{hyperref}  
\hypersetup{colorlinks=true,urlcolor=blue,citecolor=blue,linkcolor=blue}   
\usepackage{scrextend}
  {\begin{list}{}%
          {\setlength{\leftmargin}{.2cm}}%
          {\setlength{\rightmargin}{.2cm}}%
          \item[]%
  }
  {\end{list}}

\begin{document}

\title{Using disciplinary perspectives to refine conceptions of the "real world"}
\author{Abhilash Nair}
\affiliation{Department of Physics \& Astronomy,Michigan State University, 567 Wilson Rd., East Lansing, MI, 48824}
\author{Paul Irving}
\affiliation{Department of Physics \& Astronomy,Michigan State University, 567 Wilson Rd., East Lansing, MI, 48824}
\author{Vashti Sawtelle}
\affiliation{Lyman Briggs College, Michigan State University,919 E. Shaw Ln, East Lansing, MI, 48824}
\affiliation{Department of Physics \& Astronomy,Michigan State University, 567 Wilson Rd., East Lansing, MI, 48824}


\begin{abstract}
It has been reported that students often leave the introductory physics classroom believing that physics is less connected to the real world than when they entered. In this paper we aim to complicate that narrative by considering students' experiences in an introductory physics for the life-sciences course that leverages students’ disciplinary expertise in biology and chemistry as they learn physics. In a case study of three students, we probed the role of physics in their lives to challenge the typical interpretation of PER attitudinal and epistemological measures that aim to assess how students connect physics to their lives outside of the classroom. Although we find that our life-science students' rarely think of physics in their everyday lives, they make a variety of connections to the real world. We argue that in order to reflect students’ rich disciplinary experiences, our understanding of how students connect physics to life outside the classroom needs to be nuanced or expanded.
\end{abstract}

\maketitle

\section{Introduction}
It has been stated, in PER, that students often demonstrate an unfavorable shift in attitudinal measures in general and that students leave the physics classroom stating that physics is less connected to the world than when they started the course \cite{CLASS,Redish}. Items that assess the connection of physics to students’ lives appear across several attitudinal and epistemological measures such as the Views About Science Survey (VASS), Epistemological Beliefs Assessment for Physical Sciences (EBAPS), Colorado Learning Attitudes about Science Survey (CLASS), and the Maryland Physics Expectations Survey (MPEX) \cite{CLASS,Redish,VASS,EBAPS}. Students' unfavorable responses to items on these measures are often interpreted as students not perceiving the relevance of physics to the real-world, to everyday life, personal interests, or future careers \cite{CLASS,Redish,VASS}. This paper aims to complicate this oft-stated finding by exploring the types of connections students are able to make between physics and their lives outside the physics classroom.

\section{Theoretical Framework}
We posit that the clusters and items in the VASS, EBAPS, CLASS, and MPEX that assess how connected physics is to students' lives are actually part of a broader question of "how relevant is physics?" Relevance as a construct has been difficult to define \cite{Bookstein, Newton}. One of the challenges with defining relevance is  the variety of colloquial meanings -  such as significance, importance, applicability, and more \cite{Newton}. Newton explores relevance in the context of teaching science to primary and secondary students, specifically the challenges teachers face when responding to calls for more relevant science teaching \cite{Newton}. Bookstein approaches relevance in the context of information sciences, and attempts to operationalize relevance for the purposes of information retrieval \cite{Bookstein}. Building on their work, this paper operationalizes relevance in physics education to have the following characteristics:
%
%
\begin{addmargin}[.41cm]{.41cm}
1. It is a relation between the student and physics\\
2. It addresses some need, aspiration, or expectation
\end{addmargin}

To evaluate the relevance of physics we can look to see how physics is situated in the larger context of students’ lives. With an operational definition of relevance to work with, we turn to PER measures to explore which areas of students’ lives are being probed for connections to physics. We identify the following clusters as pertaining to relevance:
\begin{addmargin}[.41cm]{.41cm}
\textbf{CLASS}: Personal Interest \& Real-World Connection \\
\textbf{MPEX}: Reality Link \\
\textbf{EBAPS}: Real-life Applicability \\
\textbf{VASS}: Personal Relevance \footnote{Present in version P204, but removed in P05.07, we include it because it adds valuable insight into how questions around relevance are being asked in PER}
\end{addmargin}
In these clusters we see 2 broad types of items, those that probe (1) the connection between physics to life outside the physics classroom and (2) students' affect, motivation, and interest in physics. In this paper, we narrow our focus to connections between physics and life outside the physics classroom. Adams et. al. have reported the distinction between “whether students think that physics describes the real world and whether they actually care or think about the physics they experience in everyday life \cite{CLASS}.” We argue that for our life-science students with rich disciplinary experiences, this distinction into two categories may be insufficient to capture the diverse ways in which students are able to connect physics to their lives. We will explore how our life-science students articulate physics connections to their world  and show the complicating nuances in how these connections manifest in students’ lives.

\section{Methods}

This work is situated in the first semester of an introductory physics for the life-sciences  course that is taught in the studio format. We solicited volunteers for interviews and cross-matched using a survey to identify students who were feeling fearful, had a strong disciplinary identity,  were feeling skeptical of the course being meaningful to them, or were optimistic the course could be relevant. This paper represents work from a case study of 3 students, together they demonstrate a variety of ways in which students may connect physics to their lives. We draw from two semi-structured interviews with each of these students. The interviews included items intended to probe the relevance of physics to students’ lives. The first interview took place in the early weeks of the course, and the second interview took place near the midpoint of the semester.

\section{"Beverly"}

Beverly is a human biology major on a pre-medical track who also has volunteered her time at a clinic and the Red Cross club on campus. At the time of the study, Beverly has had physics in high school, which was a negative experience for her. 
\begin{addmargin}[.41cm]{.41cm}
\textit{“I took it [physics] senior year of high school and I hated it. My teacher was awful...I didn’t really learn anything from him”} [Int 1]
\end{addmargin}
In the early weeks of the course, Beverly doesn’t believe physics is going to be relevant to her intended future as a physician. 
\begin{addmargin}[.41cm]{.41cm}
\textit{“I don't need physics. It's not really a key aspect in my future or I don't believe it will be...I view it as more of a requirement, I just have to go through it...I talked to like various physicians I’ve shadowed...they’re like ‘I don’t really use that much physics’”} [Int 1]
\end{addmargin}
In spite of her belief that physics will not play a role in her intended future, the design of the course was such that there were multiple places in her life where physics could have been meaningful; the course explicitly leverages students’ expertise of biology and chemistry as they learn physics \cite{NEXUS}. For example, Beverly had volunteered in a wound clinic; in class, students study the motion of bacteria and neutrophils in the context of wound healing. The students were asked to use video tracking software to measure the speeds of E. coli, neutrophils, and tissue healing to determine if antibiotics needed to be prescribed. Beverly didn’t find this activity to be an authentic application of how physicians would make decisions around prescribing antibiotics. 
\begin{addmargin}[.41cm]{.41cm}
\textit{"Not too much...I actually shadowed an infectious disease doctor a couple times, and by the time his patients got to him they needed the antibiotics, it wasn't just a matter of if they would need them..."} [Int 1]
\end{addmargin}
Beverly also  recalls conversations with her family around the dangers of her driving a small car and the implications of a collision; in class, students modeled a collision between a car and an SUV. Beverly states that it’s clear which car does better, and doesn’t find the unpacking of physics laws to be relevant to her future. 
\begin{addmargin}[.41cm]{.41cm}
\textit{“Most of us could figure out if it’s an SUV vs. a small car in a collision, most of us could pretty much guess which vehicle would do worse in that situation...the bigger object tends to fair a little better”} [Int 1]
\end{addmargin}
%
%
In the second interview Beverly revisits the car crash, and states that she can better explain what happens in collisions, but doesn't label it as thinking of physics concepts.  This is consistent with Beverly's previous statements that she finds physics reasoning in car collisions to be intuitive.
%
%
%
\begin{addmargin}[.41cm]{.41cm}
\textbf{I}: \textit{Do you think think activities like this one [investigating collisions lab] have equipped you to answer questions outside of class?}
\\
\textbf{Beverly}: \textit{...it was actually weird my cousin was just in a car crash a week ago... I could like pair that with this knowledge and be like OK so nothing too awful could have happened, she had airbags which we've looked into a little bit [in class]. Like I felt like a little more secure in that knowledge}
\\
\textbf{I}: \textit{You found yourself thinking of physics concepts when that happened or?}
\\
\textbf{Beverly}: \textit{Slightly...I guess I've always thought about them, I just didn't define them or label them as physics.} [Int 2]
\end{addmargin}
Beverly recognizes that physics can be applied to real world situations, but states  that she won’t need the detailed physics content knowledge. Beverly's unfavorable responses connecting physics to the real world could be interpreted as Beverly believing that “ideas learned in physics have little relation to experiences outside the classroom” \cite{Redish}. We contend that Beverly’s responses are more nuanced than this interpretation allows. One issue is the difference between Beverly recognizing that physics can be used in the real world and her believing that she doesn’t need to bring in physics in those experiences. Beverly's statements are about the practical necessity of using physics in these real world scenarios, and not a reflection of how much she values thinking about physics. These statements reflect a sophistication that's grounded in her medical experience as well as her sense for when physics is \textit{needed}. Although Beverly is able to connect physics to real world events, she insists that this is something she has always done and doesn't label it as physics. This is consistent with earlier statements that she doesn't think of physics in everyday life.
%
%
%
%
%
%
%
\begin{addmargin}[.41cm]{.41cm}
\textbf{I}: \textit{Do you think of physics outside the classroom?}
\\
\textbf{Beverly}:  \textit{...I don’t really do. If I throw something, I don’t really think about it} [Int 1]
\end{addmargin}
The notion that Beverly doesn’t typically think of physics outside the physics classroom is not surprising taking into account her previous experiences and her statements that she doesn’t \textit{need} physics.  

\section{"Maria"}

Maria is a microbiology major with a minor in epidemiology who identifies strongly as a microbiologist. She has leadership positions in multiple biology related organizations and works in a microbiology research lab. Maria recalls her high school physics experience as being disconnected from her interests. 
\begin{addmargin}[.41cm]{.41cm}
\textit{"The last time I’ve had physics was sophomore year of high school...I just don’t think they did a very good job of connecting it back to everyone's interests...it was just theoretical pure physics. so, not my thing..."} [Int 1]
\end{addmargin}
Maria is optimistic that this course may be more relevant. 
\begin{addmargin}[.41cm]{.41cm}
\textit{"I think in this course it may be a little more [relevant], 'cause of the [biology] connections, otherwise I would probably say no [laughs]"} [Int 1]
\end{addmargin}
For Maria, the wound healing activity was meaningful and the type of connection she was hopeful for. 
\begin{addmargin}[.41cm]{.41cm}
\textit{“I kind of knew like...it was hopefully going to be like this...I'm a microbiology major, I know physics is important for what I want to do but like physics like I was taught in high school...it's too conceptual, too theoretical, but this [wound healing activity] was you know like here's where you would apply it, like what concepts to use specifically in microbiology.”} [Int 1]
\end{addmargin}
%
%
When asked if she is finding physics in any of her other courses, Maria readily sees areas in her courses where physics may play a role. 
\begin{addmargin}[.41cm]{.41cm}
\textit{“My past epi [epidemiology] course was really focused on like...osteoarthritis...I can definitely see how you would be like focused on like the physics of it all, like what causes the fractures, what can we do to prevent them and things like that. Not so far in like prokaryotic physio, but I'm feeling once we get to flagella and pillae and things that are moving, maybe a little more.”} [Int 1]
\end{addmargin}
This connection to her prokaryotic physiology course ends up foreshadowing a strong, meaningful connection to physics for Maria midway through the semester.
\begin{addmargin}[.41cm]{.41cm}
\textbf{I}: \textit{So in microbio, if you have flagella like that, would physics help you answer that?}
\\
\textbf{Maria}: \textit{Oh for sure. [laughs]...you have the chemistry interactions that get you the movement, they happen inside the cell, and then the physics can explain those chemistry interactions} [Int 2]
\end{addmargin}
When asked how she realized physics can explain why, Maria explains that it didn’t happen until she had the tools to make the connection between microbiology and physics. 
\begin{addmargin}[.41cm]{.41cm}
\textit{"...sure you can see something and be like ‘physics probably explains that’ but I don’t know physics, so why would I think about it that way if I don’t have that tool?...taking this course, the microbio course alongside physics where things like work and torque and force are coming up in a field that I know about, it helps you see.”} [Int 2]
\end{addmargin}
%
%
This moment is so powerful for Maria that she suggests these types of investigations is the physics she would imagine herself doing. 
\begin{addmargin}[.41cm]{.41cm}
\textit{“The physics I'd be interested in doing is something like this [paramecium activity]...it’s too late in my career, my parents would kill me if I switched my major... I'm interested in how it [physics] relates to the macroscopic biological world. I know other people do other things, but this is where the physics I like.”} [Int 2]
\end{addmargin}
%
%
Maria’s real-world connections are shaped by her disciplinary identity as a microbiologist. When physics can connect to the world she identifies with, we see Maria articulate connections linking physics content knowledge with biology. When asked if she thinks of physics outside of the classroom, Maria responds 
\begin{addmargin}[.41cm]{.41cm}
\textit{“...not so much”} [Int 1]
\end{addmargin}
Similarly, Maria states she doesn’t talk about physics with friends or family.
\begin{addmargin}[.41cm]{.41cm}
\textbf{I}:  \textit{Do you talk to your family or friends about physics?}
\\
\textbf{Maria}: \textit{No [shakes head] other than my friends that are in the class I don't know who.} [Int 1]
\end{addmargin}
Maria's connections to physics are different in nature to Beverly's. Beverly's connections involve her projecting forward to her intended future career as a physician and the practices of a physician, informed by observing and talking with physicians. Maria, on the other hand, makes connections across the disciplines, informed by her strong identification as a microbiologist. Overall we see the intersection of physics and biology aligning with Maria’s interests.
%
%
%
%

\section{"Miles"}

Miles is a first-generation college student majoring in biochemistry with a minor in bioethics who has never taken a physics course. He enjoys biology and conducts research in a biochemistry lab. Miles is fearful of physics and recalls horror stories. 
\begin{addmargin}[.41cm]{.41cm}
\textit{“I’m most nervous about physics to be completely honest, I've never taken physics before, never...you hear horror stories about physics”} [Int 1]
\end{addmargin}
%
%
From his experience as a tutor for the biology and chemistry portions of the MCAT, he has seen negative experiences others have had with physics 
\begin{addmargin}[.41cm]{.41cm}
\textit{“That was kind of like the first time I saw physics like right up close...tutors were pulling their hair out, these kids just looked terrible after the physics parts...”} [Int 1]
\end{addmargin}
%
%
This was Miles’s first time learning physics and he was not certain how physics fit into the larger world but he repeatedly states that \textit{“it has to be physics”} underlying many of the phenomena of the world. Miles uses the context of water molecules moving away from each other to explain how he sees biology, chemistry, and physics as being related \begin{addmargin}[.41cm]{.41cm}
\textit{“There's obviously driving forces behind it like I said I never really took physics so I didn't know...it has to be physics, it has to be. I think physics is a driving force behind everything, it has to be.”} [Int 1]
\end{addmargin}
%
%
When asked if he sees physics concepts in other courses, Miles points to topics in his chemistry course. 
\begin{addmargin}[.41cm]{.41cm}
\textit{“Yeah I’ve seen glimpses, and I just, it has to be physics, like it has to be physics. In my mind it has to be physics but I just don’t know like how exactly yet.”} [Int 1]
\end{addmargin}
%
%
Miles looks to make connections between physics and his other coursework. When asked if he still has open questions about diffusion, he is interested in knowing why diffusion occurs. 
\begin{addmargin}[.41cm]{.41cm}
\textit{“I mean I guess like why it [diffusion] happens. I mean I know why, I can say because of entropy because you discretely, it's favorable to increase disorder. But I guess I don’t know why exactly that's a thing, and I feel like that's why physics comes into play because that's the driving force behind all this movement, I think?”} [Int 1]
\end{addmargin}
Early in the course, Miles was uncertain where physics fit into the larger picture, but was certain physics had a role to play. After the unit of diffusion, in Interview 2, he is finding physics as connecting to biological processes.
%
%
\begin{addmargin}[.41cm]{.41cm}
\textbf{I}: \textit{Do you think the idea of diffusion connects to biological molecules?}
\\
\textbf{Miles}: \textit{Oh yeah, for sure yeah yeah...I think that's relevant in any type of like physiology or biochemistry...things float around in solution for a reason, things are kept at different concentrations for a reason...} [Int 2]
\end{addmargin}
Miles’s connections to the real world are more general than Maria’s specific disciplinary connections to biology and different from Beverly’s connections to car crashes and medical care. He believes that physics is important and expresses a sense that physics is the discipline underlying most phenomena. Similar to what we see with Beverly and Maria, Miles does not think of physics outside the classroom and does not talk with friends or family about physics. This is not surprising as this is his first physics course; he is still forming his conception of what physics \textit{is}.
\begin{addmargin}[.41cm]{.41cm}
\textbf{I}: \textit{Do you think of physics outside of the classroom?}
%
%
\\
\textbf{Miles}: \textit{No, literally never...} 
\\
\textbf{I}: \textit{Do you [talk to] family or friends about physics?}
\\
\textbf{Miles}: \textit{Not once.}
\\
\textbf{I}: \textit{Or even about the concepts you learned in class recently?}
\\
\textbf{Miles}: \textit{No, [laughs] no.} [Int 1]
\end{addmargin}

\section{Discussion and Conclusion}
%
%

In this paper we explored how PER surveys have aligned with an operationalized construct of relevance in physics education, especially in the ways they probe the relation between physics and students' lives. Beverly, Maria, and Miles do not commonly think of physics outside the classroom or talk about physics in their everyday life. Our larger data set suggests that thinking of or talking about physics in everyday life is rare for life-science students. We don’t find this result surprising, the expectation that non-physics major students should perceive connections to physics in their everyday life is optimistic. However, our students have a rich set of disciplinary experiences. The separation of their connections to the real world into only two categories \cite{CLASS} is insufficient to capture the diverse ways students connect physics to their world.
%
%

Beverly, Maria, and Miles are able to connect to the real world, but in different, nuanced ways. Beverly understands the role physics \textit{can} play, but generally finds bringing in of physics to be (1) inauthentic based on her medical experience and (2) unnecessary to understand events like car collisions or throwing of a ball. Maria is able to make specific cross-disciplinary connections between physics and biology. She finds these connections engaging and discovers a new interest she was not aware of. Miles attributes the underlying mechanism for most things to physics. As he experiences more physics, he believes he will be able to more strongly articulate why physics is relevant to his disciplinary interests in biology and chemistry.
%
%
%
%
%
%
%
%
%
%
%
%

In this paper, our goal is to complicate the interpretation of unfavorable shifts in attitudinal and epistemological measures. We argue that students do see the relevance of physics in their lives, but make sophisticated judgments on the role that physics plays. Our cases provide evidence that thinking of physics in everyday life manifests differently in student experiences and that students often make these connections without valuing them, which adds a complexity to interpreting results from these measures.
%
%
%
%
%
%

The relevance of physics to a student’s life can take on many forms, and the collection of PER measures probes some of those ways. The cluster names are often descriptive labels put on a set of items that have been found to align together by validating the survey. Suggesting that a set of students respond unfavorably in connecting physics to the real world is a significant and meaningful statement. Our case studies demonstrate that the space around students' connections to the real world is complex and that measures of these connections should be nuanced or expanded if they are to reflect students' rich disciplinary experiences. This becomes especially important when teaching a physics for the life-sciences course in which we actively work to make physics relevant to life outside the physics classroom.
\acknowledgments{We want to thank the PERL@MSU group for helpful discussions. This work is supported by Lyman Briggs College and the physics department at MSU.}


\begin{thebibliography}{1}%
\makeatletter
\providecommand \@ifxundefined [1]{%
 \@ifx{#1\undefined}
}%
\providecommand \@ifnum [1]{%
 \ifnum #1\expandafter \@firstoftwo
 \else \expandafter \@secondoftwo
 \fi
}%
\providecommand \@ifx [1]{%
 \ifx #1\expandafter \@firstoftwo
 \else \expandafter \@secondoftwo
 \fi
}%
\providecommand \natexlab [1]{#1}%
\providecommand \enquote  [1]{``#1''}%
\providecommand \bibnamefont  [1]{#1}%
\providecommand \bibfnamefont [1]{#1}%
\providecommand \citenamefont [1]{#1}%
\providecommand \href@noop [0]{\@secondoftwo}%
\providecommand \href [0]{\begingroup \@sanitize@url \@href}%
\providecommand \@href[1]{\@@startlink{#1}\@@href}%
\providecommand \@@href[1]{\endgroup#1\@@endlink}%
\providecommand \@sanitize@url [0]{\catcode `\\12\catcode `\$12\catcode
  `\&12\catcode `\#12\catcode `\^12\catcode `\_12\catcode `\%12\relax}%
\providecommand \@@startlink[1]{}%
\providecommand \@@endlink[0]{}%
\providecommand \url  [0]{\begingroup\@sanitize@url \@url }%
\providecommand \@url [1]{\endgroup\@href {#1}{\urlprefix }}%
\providecommand \urlprefix  [0]{URL }%
\providecommand \Eprint [0]{\href }%
\providecommand \doibase [0]{http://dx.doi.org/}%
\providecommand \selectlanguage [0]{\@gobble}%
\providecommand \bibinfo  [0]{\@secondoftwo}%
\providecommand \bibfield  [0]{\@secondoftwo}%
\providecommand \translation [1]{[#1]}%
\providecommand \BibitemOpen [0]{}%
\providecommand \bibitemStop [0]{}%
\providecommand \bibitemNoStop [0]{.\EOS\space}%
\providecommand \EOS [0]{\spacefactor3000\relax}%
\providecommand \BibitemShut  [1]{\csname bibitem#1\endcsname}%
\let\auto@bib@innerbib\@empty
\bibitem [{Note1()}]{Note1}%
  \BibitemOpen
  \bibinfo {note} {Present in version P204, but removed in P05.07, we include
  it because it adds valuable insight into how questions around relevance are
  being asked in PER}\BibitemShut {NoStop}%
\end{thebibliography}%


\begin{thebibliography}{99}
\bibitem{CLASS} W. K. Adams et al., Phys. Rev. Spec. Top. - Phys. Educ. Res. 2, 1 (2006).
\bibitem{Redish} E. F. Redish, J. M. Saul, and R. N. Steinberg, Am. J. Phys. 66, 212 (1998).
\bibitem{VASS} I. Halloun and D. Hestenes, Sci. Educ. 7, 553 (1998).
\bibitem{EBAPS} A. Elby, J. Frederiksen, C. Schwarz, and B. White, (1998).
\bibitem{Bookstein} A. Bookstein, J. Am. Soc. Inf. Sci. 30, 269 (2007).
\bibitem{Newton} D. P. Newton, Educ. Philos. Theory 20, 7 (1988).
\bibitem{NEXUS} E. F. Redish et al., Am. J. Phys. 82, 368 (2014).
\end{thebibliography}
\end{document}